\documentclass[submission,copyright]{eptcs}
\providecommand{\event}{ICE 2012} 

\usepackage{graphics,epsfig,times}

\usepackage[latin1]{inputenc}
\usepackage[english]{babel}
\usepackage[centertags]{amsmath}
\usepackage{amsfonts}
\usepackage{amsthm}
\usepackage{amssymb}
\usepackage{newlfont}
\usepackage{amsmath}
\usepackage{latexsym}
\usepackage{mathrsfs}
\usepackage{stmaryrd}
\usepackage{color}
\usepackage{graphicx}
\usepackage{setspace}
\usepackage{subfig}
\usepackage{multirow}
\usepackage{url}

\def\SemiDom{D}

\def\Semiring{T}
\def\semA{d_{1}}
\def\semB{d_{2}}
\def\semC{d_{3}}

\newcounter{propc}
\theoremstyle{definition}
\newtheorem{defin}{Definition}
\newtheorem{ex}{Example}
\newtheorem{prop}[propc]{Property}

\theoremstyle{plain}

\newtheorem{theorem}{Theorem}

\newenvironment{example}
{\nobreak\begin{ex}}
{\qed\end{ex}\nobreak}

\newenvironment{property}
{\nobreak\begin{prop}}
{\end{prop}\nobreak}

\newenvironment{definition}
{\nobreak\begin{defin}}
{\end{defin}\nobreak}

\renewcommand{\epsilon}{\varepsilon}

\newcommand{\lreq}{$\lambda^{req}$}

\newcommand{\ttxt}[1]{\ensuremath{\mathrm{\mathtt{#1}}}}

\newcommand{\osstep}[3]{\ensuremath{\left\langle {#1} \right\rangle \rightarrow_{#2} \left\langle {#3} \right\rangle}}
\newcommand{\tosstep}[3]{\ensuremath{\left\langle {#1} \right\rangle \rightarrow^\ast_{#2} \left\langle {#3} \right\rangle}}

\newcommand{\cvd}{\begin{flushright}$\blacksquare$\end{flushright}}

\newcommand{\true}{\mathit{tt}}
\newcommand{\false}{\mathit{ff}}

\newcommand{\sem}[3]{\mbox{\ensuremath{\llbracket {#1}\rrbracket_{#3}}}}
\newcommand{\bind}[2]{^{#2}/_{#1}}

\newcommand{\irule}[2]{\frac{\,\textstyle\rule[-1.3ex]{0cm}{3ex}#1\,}%
{\textstyle\rule[-.5ex]{0cm}{3ex}#2}}

\newcommand{\htype}[3]{\mbox{\ensuremath{#1 \xrightarrow{#2} #3}}}

\newcommand{\orsyn}{\ensuremath{\,\mid\,}}

\newcommand{\hbra}{
\hbox to \textwidth{\vrule width0.3mm height 1.8mm depth-0.3mm
                    \leaders\hrule height1.8mm depth-1.5mm\hfill
                    \vrule width0.3mm height 1.8mm depth-0.3mm}}
\newcommand{\hket}{
\hbox to \textwidth{\vrule width0.3mm height1.5mm
                    \leaders\hrule height0.3mm\hfill
                    \vrule width0.3mm height1.5mm}}

\newcommand{\defeq}{\stackrel{def}{=}}
\renewcommand{\quote}[1]{\lq\lq{#1}\rq\rq}

\newcommand{\trel}[3]{#1 \vdash #3}

\newcommand{\Srv}{\mathrm{\mathtt{Srv}}}

\newcommand{\req}[2]{\textup{\texttt{req}}_{#1} \hspace{2pt} {#2}}

\newcommand{\ifthen}[3]{\textup{\texttt{if}}\hspace{2pt}#1\hspace{2pt}\textup{\texttt{then}}\hspace{2pt}#2 \hspace{2pt} \textup{\texttt{else}} \hspace{2pt} #3}
\newcommand{\ifc}{\textup{\texttt{if}}\hspace{2pt}}
\newcommand{\thenc}{\hspace{2pt}\textup{\texttt{then}}\hspace{2pt}}
\newcommand{\elsec}{\hspace{2pt}\textup{\texttt{else}}\hspace{2pt}}

\newcommand{\sfrm}[2]{\mbox{\ensuremath{#1\!\left[#2\right]}}}

\newcommand{\labs}[3]{\mbox{\ensuremath{\lambda_{#1}{#2}.{#3}}}}
\newcommand{\mfrm}[2]{\mbox{\ensuremath{#1\left\langle #2 \right\rangle }}}

\newcommand{\fork}[2]{\textup{\texttt{fork}}\hspace{2pt}{#1}\hspace{2pt}\textup{\texttt{and}}\hspace{2pt}{#2}}
\newcommand{\forkc}{\textup{\texttt{fork}}\hspace{2pt}}
\newcommand{\andc}{\hspace{2pt}\textup{\texttt{and}}\hspace{2pt}}

\newcommand{\mann}[2]{{#1} \# {#2}}

\newcommand{\unit}[0]{\mathit{unit}}

\newcommand{\Risk}{\ensuremath{\mathbf{Risk}} }

\begin{document}
\title{Metric-Aware Secure Service Orchestration\thanks{The research leading to these results has received funding from the European Union Seventh Framework Programme (FP7/2007-2013) under grant numbers 257930 (ANIKETOS), 256980 (NESSOS) and 257876 (SPACIOS).}}


\author{Gabriele Costa
\institute{Dipartimento di Informatica, Sistemistica e Telematica\\
Universit\`a di Genova}
\email{gabriele.costa@unige.it}
\and
Fabio Martinelli
\institute{Istituto di Informatica e Telematica\\
Consiglio Nazionale delle Ricerche}
\email{fabio.martinelli@iit.cnr.it}
\and
Artsiom Yautsiukhin
\institute{Istituto di Informatica e Telematica\\
Consiglio Nazionale delle Ricerche}
\email{artsiom.yautsiukhin@iit.cnr.it}
}
\def\titlerunning{Metric-Aware Secure Service Orchestration}
\def\authorrunning{G. Costa, F. Martinelli \& A. Yautsiukhin}

\maketitle

\begin{abstract}
Secure orchestration is an important concern in the internet of service. Next to providing the required functionality the composite services must also provide a reasonable level of security in order to protect sensitive data. Thus, the orchestrator has a need to check whether the complex service is able to satisfy certain properties. Some properties are expressed with metrics for precise definition of requirements. Thus, the problem is to analyse the values of metrics for a complex business process.

In this paper we extend our previous work on analysis of secure orchestration with quantifiable properties. We show how to define, verify and enforce quantitative security requirements in one framework with other security properties. The proposed approach should help to select the most suitable service architecture and guarantee fulfilment of the declared security requirements.
\end{abstract}

%

\addtolength{\textheight}{.4in}
\section{Introduction}
\label{sec:Introduction}

Orchestration of complex web services is a multidimensional problem.
Various criteria must be considered when different alternatives exist.
Typically, one of such criteria is \emph{security}.
Recently, the security issues of service composition are receiving major attention~\cite{NIEL-07-CSF,ROSS-09-ICUMT,Bartoletti05csfw,Bravetti09contract,Padovani08contract,Castagna09contracts}.
Among them, formal methods have been successfully applied for modelling and analysing several different aspects of service security.
In practice, these techniques generate a formal abstraction of the services under analysis.
Then, a verification procedure is applied to find a formal proof of compliance between the model and the security specifications.

The first difficulty arises from service abstraction.
Indeed, it is crucial that services are modelled in a \quote{safe} way, i.e., without neglecting any security-relevant behaviour they can generate.
The problem is that this feature is not always guaranteed as specification and implementation is often developed independently.

Although several, effective algorithms for software verification exist, e.g., model checking~\cite{Clarke86mc}, they often require some modification to be applied to web services.
Indeed, the algorithms typically check the compliance between a specification and a model and, if the check fails, they return a description of the detected error, e.g., a behaviour of the model that violates the specification.
However, web services are designed and developed separately and they commonly have different and independent security requirements.
Moreover, they are oriented to the composition and they can produce many different models, i.e., one for each possible orchestration.
Hence, the verification process cannot just focus on 
an illegal orchestration, but should help in finding valid ones.

Service usages are often based on \emph{security metrics}.
Metrics conveniently use mathematical values to represent some \quote{qualities} of a service.
Several authors, e.g., see~\cite{YU-05-EEE,MASS-07-QoP}, proposed mathematical models for the definition and composition of security metrics.

In this paper we propose an extension of previous work (see~\cite{Costa10modular,Costa11jsa}) on secure service orchestration integrating facilities for composing and verifying security metrics.
In particular, we start from the service model proposed by Bartoletti et al.~\cite{Bartoletti05csfw}.
Roughly, they propose a type and effect system for producing safe abstractions of the behaviour of web services.
Then, the authors verify these abstractions against the security policies, locally specified by each service, to find a valid composition.

We extend their model by introducing metric checks and metric annotations on their abstractions. We use a mathematical structure, called c-semiring, in order to generalise our model and be independent from the metrics used for the analysis, but still be able to reason on these metrics. Metric annotations are obtained through a new, improved type and effect system.
In this way, we generate metric-annotated abstractions which contain both security and metric requirements.
All the requirements are applied to different portions of the service orchestration through a local scope.

The main advantage of this approach is the possibility to model and compose both security and metric requirements in a single framework.
Service developers apply security policies and metric checks to some parts of their services.
Our type and effect system extracts \emph{history expressions} from the implementation of the services.
History expressions safely denote the behaviour of service invocations.
Within a history expression, the type and effect system adds extra annotations for metrics, metric checks and security framings.
Then, we adopt the same verification procedure described in~\cite{Bartoletti05csfw} with special pre-processing steps for assigning correct metric labels to each service.
The final result is a complete framework for defining, modelling, verifying, and enforcing both security and metric requirements in order to find valid service orchestrations.

This paper is structured as follows.
Section~\ref{sec:example} introduces the working example we will develop during our presentation.
In Section~\ref{sec:sstructure} we describe our extension of the programming language~\lreq~and we define its operational semantics.
Then, Section~\ref{sec:types} presents our type and effect system and Section~\ref{sec:analysis} describes the analysis of security and metric requirements.
Finally, Section~\ref{sec:conclusion} concludes the paper.

\section{Running example}\label{sec:example}

The travel agency BestTravel offers a travel planning service to its customers.
BestTravel exploits existing services for implementing the complex task of $(i)$ booking a connection (consisting of one or more flights) to the destination, $(ii)$ booking a hotel room, $(iii)$ paying the acquired items (i.e., flights and hotel room), and $(iv)$ providing the customer with a signed receipt.
As usual in service-oriented architectures, the four subtask described above are provided by existing web services.

\begin{figure}[ht]
\begin{center}
  \includegraphics[width=400pt]{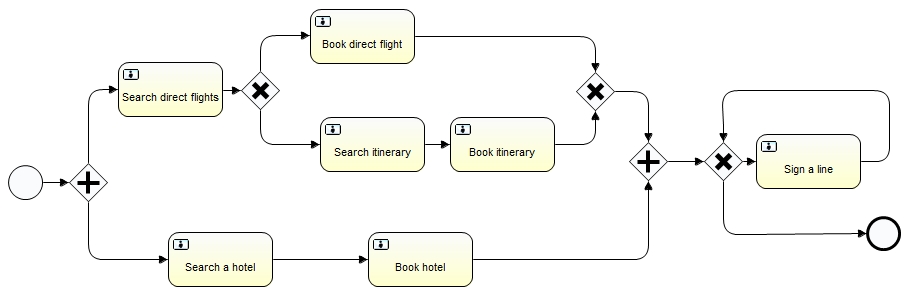}
  \caption{Abstract workflow for BestTravel.}
    \label{fig:expWorkflow}
\end{center}
\end{figure}

The service developer starts from an abstract workflow describing the behaviour of BestTravel and  produces a corresponding implementation.
The abstract workflow depicts the atomic operations that the service must implement and how they compose each other.
In the case of BestTravel, most of the atomic operations are invocations to other services.
Figure~\ref{fig:expWorkflow} shows the abstract workflow of BestTravel.

Reading Figure~\ref{fig:expWorkflow} (from left to right), we can understand the service behaviour.
In words, a session of BestTravel works as follows.
The service runs two procedures in parallel (rooted in \rotatebox{45}{$\boxtimes$}).
The first one (upper path of the workflow) is responsible for booking a flight connection for the travel destination.
In practice, BestTravel invokes a service looking for a direct flight, i.e., \emph{search direct flight}.
Then the execution can take two alternative branches (\rotatebox[origin=c]{45}{$\boxplus$} node): it can invoke a payment service for booking the flight, i.e., \emph{book flight}, or it can start a new research for a multiple-flight connection, namely an \emph{itinerary}, and book it, i.e., \emph{search itinerary} and \emph{book itinerary}.
Concurrently, the second process (lower path) invokes services for searching and booking a hotel, i.e., \emph{search hotel} and \emph{book hotel}.
When the two parallel procedures terminate, BestTravel iteratively invokes a digital signature service, i.e., \emph{sign line}, for applying integrity and authenticity tokens to the hotel receipt and terminates.

A requirement of BestTravel is to have \emph{risk} level of the performed tasks (in particular, flight booking, hotel reservation and receipt signature) less than 75. Therefore, two problems must be solved: (i) \emph{statically} estimate risk for the composition plans; (2) in case some execution path in the composition plan fails the requirement, \emph{dynamically} check the risk of selected paths and prevent the failure of the requirement if a risky path is selected.

\section{Service structure}\label{sec:sstructure}

In this section we present an extended version of $\lambda$-calculus, called~\lreq~\cite{Bartoletti05csfw}.
First, we extend our previous work with two main novelties: \emph{parallel composition} and \emph{metric facilities}.
Parallel agents in this work are defined without modifying the original syntax of the calculus.
We obtain it by re-defining the operational semantics of~\lreq.
Second, we incorporate metrics into our formalism using special operations for denoting metric annotations and metric constraints.
These operators are interpreted in a c-semiring mathematical structure.
Metric facilities allow us to model metrics which are used in service composition.

\subsection{Syntax}
\label{sec:Syntax}

\begin{table}[t]
\centering
\begin{minipage}{\columnwidth}
\hbra
\begin{footnotesize}
\begin{tabular}{l | l}
\begin{minipage}{0.49\columnwidth}
\noindent
\begin{tabular}{r l @{\hspace{12pt}} l}
 $e,e'$ & \textnormal{::=} \\ 
 & $\ast$ & unit \\
 & $r$ & resource \\
 & $x$ & variable \\
 & $\alpha(e)$ & access event \\
 & \ifthen{$b$}{$e$}{$e'$} & branch \\
 & \labs{z}{x}{e} & abstraction \\
 & $e\,e'$ & application \\
 & \sfrm{\varphi}{e} & security framing \\
 & \mfrm{\gamma}{e} & metric framing \\
 & $\req{\rho}{\htype{\tau}{}{\tau'}}$ & service request \\ [0.5ex]
\end{tabular}
\end{minipage}
&
\begin{minipage}{0.48\columnwidth}
\[
\labs{}{x}{e} \defeq \labs{z}{x}{e} \;\textnormal{with }z \not\in fv(e)
\]
\[
\labs{}{}{e} \defeq \labs{}{x}{e} \;\textnormal{with }x \not\in fv(e)
\]
\[
e;e' \defeq (\labs{}{}{e'})e \qquad \fork{e}{e'} \defeq (e';\labs{}{x}{x})e
\]
\[
(\req{\rho}{\htype{\tau}{\varphi,\gamma}{\tau'}})e \defeq \sfrm{\varphi}{\mfrm{\gamma}{(\req{\rho}{\htype{\tau}{}{\tau'}})e}}
\]
where $fv$ is the standard function returning the set of free variables of an expression $e$.
\end{minipage}
\end{tabular}
\\
\hket
\end{footnotesize}
\end{minipage}
\caption{Syntax of \lreq and abbreviations}\label{tab:syntax}
\end{table}

First, we define the syntax of expressions $e,e'$ as shown in Table~\ref{tab:syntax}.
Briefly, $\ast$ is the closed, side effects-free expression, $r,r' \in \mathcal{R}$ denotes system resources and $x,y$ are variables.
Access events $\alpha(e), \beta(e')$ represent the access to a certain resource, resulting from the evaluation of the event argument, through a specific operation/channel (e.g., $\alpha$ and $\beta$).
Conditional term \ifthen{$b$}{$e$}{$e'$} represents a branch between two expressions (where $b$ is a boolean guard).
A function is defined through the term \labs{z}{x}{e}, where $e$ is the function body in which $x$ is the formal parameter and $z$ denotes the function itself (for recursive invocations).
Instead, the term $e\,e'$ denotes the application of a function $e$ to a parameter $e'$. We feel free to use parenthesis for grouping either a function or its argument in order to improve readability.
Security framing is used to apply the scope of a security policy $\varphi$ to a term.
We also use metric framing for expressing a term laying in the scope of a metric constraint $\gamma$.
Finally, a service request $\req{\rho}{\htype{\tau}{}{\tau'}}$ denotes the invocation of a service having a certain functional interface, i.e., $\htype{\tau}{}{\tau'}$ shows that the function requires a type $\tau$ as input and produces type $\tau'$ as output, and is labelled with a unique identifier $\rho$. Although, it is hard to create the \lreq~representation for non experts such the model may be created automatically, similar to transformation of Java code \cite{BART-09-ENTCS}.

For the sake of presentation, we introduce some useful abbreviations (see Table~\ref{tab:syntax}).
%
%
%
Moreover, to improve the readability we feel free to use simple expressions for conditional guards, e.g., \ttxt{is\_available} or \ttxt{is\_empty}, which have a straightforward interpretation in the context we use them.
We also use upper cases for resources, e.g.,~\ttxt{HOTEL} and \ttxt{FLIGHT}, and lower cases for actions, e.g., \ttxt{book}($\ldots$) and \ttxt{buy}($\ldots$).
%

According to the standard~\lreq~theory, we define security policies through \emph{usage automata}~\cite{Bartoletti09ua}.
Usage automata resemble non deterministic finite state automata (NFA) defined over the alphabet of access events.
A sequence of actions is compliant with a certain policy if its corresponding usage automata does not reach a final, offending state reading the trace, i.e., valid traces are those rejected by the automata (see \cite{Bartoletti09ua} for details).

Our main focus in this section is on the definition of metric constraints.
Indeed, we introduce a syntax for defining metric checks which then we apply through metric framing.
In particular a metric check has the form $\gamma = T \geq_T d$ where $T$ is a metric name, $\geq_T$ is its order relation and $d$ is an element of $T$. Here we slightly abuse our notation for the sake of simplicity, in order to show that the metric computed for a business process must be better than some predefined value (i.e., threshold). In practice, a metric check is satisfied by a value $d'$ if $d' \geq_T d$. If so we write $d' \in \gamma$.

%

\begin{example}\label{ex:serviceimpl}
We continue our running example.
We assume the (sets of) resources:
$\mathcal{I} = \{\ttxt{ITINERARY}\}$, $\mathcal{F} = \{\ttxt{FLIGHT\_No}, \ttxt{NO\_FLIGHT}\}$, $\mathcal{H} = \{\ttxt{HOTEL\_RESV}\}$, $\mathcal{B} = \mathcal{I} \cup \mathcal{F} \cup \mathcal{H}$ and $\mathcal{D} = \{\ttxt{RCPT}, \ttxt{SIGNED\_DOC}\}$.
In Figure~\ref{fig:serviceimpl} we propose the \lreq~implementation of the services informally introduced in Section~\ref{sec:example}.

\begin{figure}
\begin{tabular}{l  l}
\begin{minipage}{0.48\columnwidth}
\noindent
\begin{footnotesize}
\begin{tabular}{| l | l }
1 & 
  \labs{}{x}{(\ttxt{search\_flight\_for}(x);} \\
           & $\quad$\ifc\ttxt{is\_available} \\
           & $\qquad$\thenc\ttxt{reserve}(\ttxt{FLIGHT\_No});\ttxt{FLIGHT\_No} \\
           & $\qquad$\elsec\ttxt{NO\_FLIGHT}) \\ [1.5ex]

2 & 
  \labs{}{x}{(\ttxt{search\_flight\_for}(x);} \\
           & $\quad$\ifc{\ttxt{is\_available}} \\
           & $\qquad$\thenc{\ttxt{reserve}(\ttxt{FLIGHT\_No});\ttxt{FLIGHT\_No}} \\
           & $\qquad$\elsec{\ifc{\ttxt{can\_overbook}}} \\
           & $\qquad\quad$\thenc{\ttxt{overbook}(\ttxt{FLIGHT\_No});\ttxt{FLIGHT\_No}} \\
           & $\qquad\quad$\elsec{{\ttxt{NO\_FLIGHT}}}) \\ [1.5ex]

3 & 
  \labs{}{x}{(\ttxt{generate\_travel\_to}(x);\ttxt{reserve}(\ttxt{ITINERARY});} \\
           & $\quad\ttxt{insurance}(\ttxt{ITINERARY});\ttxt{ITINERARY})$ \\ [1.5ex]

4 & 
  \labs{}{x}{(\ttxt{generate\_travel\_to}(x);\ttxt{reserve}(\ttxt{ITINERARY});} \\
           & $\quad\ttxt{ITINERARY})$ \\ [1.5ex]
\end{tabular}
\end{footnotesize}
\end{minipage}
&
\begin{minipage}{0.5\columnwidth}
\noindent
\begin{footnotesize}
\begin{tabular}{ | l | l }
5 & 
  $\labs{}{x}{(\ttxt{find\_hotel\_3s}(x);\ttxt{book}(\ttxt{HOTEL});}$ \\
 		   & $\quad\ttxt{HOTEL\_RESV})$ \\ \\

6 & 
  \labs{}{x}{(\ifc{\ttxt{high\_season}}} \\
			 & $\quad\thenc{\ttxt{find\_hotel\_2s}(x)}\elsec{\ttxt{find\_hotel\_4s}(x)};$ \\
			 & $\qquad\ttxt{book}(\ttxt{HOTEL});\ttxt{HOTEL\_RESV})$ \\ \\
			
7 & 
  \labs{}{x}{((\ifc{\ttxt{registered\_user}}} \\
             & $\quad\thenc{\ast}\elsec{\ttxt{var\_charge}(x)});\ttxt{buy}(x);\ttxt{RCPT})$ \\ \\

8 & 
  \labs{}{x}{(\ttxt{const\_charge}(x);\ttxt{buy}(x);\ttxt{RCPT})} \\ \\

9 & 
  \labs{}{x}{\ttxt{sign\_64}(x);\ttxt{SIGNED\_DOC}} \\ \\

10 & 
  \labs{}{x}{\ttxt{sign\_128}(x);\ttxt{SIGNED\_DOC}} \\ \\
\end{tabular}
\end{footnotesize}
\end{minipage}
\end{tabular}
\caption{Implementation of the services of Example~\ref{ex:serviceimpl}.}
\label{fig:serviceimpl}
\end{figure}

Intuitively, service 1 receives an input airport $x$ and searches a direct flight (action \ttxt{search\_flight\_for}).
Then, depending on the \ttxt{is\_available} boolean flag, the service either reserves a seat (\ttxt{reserve}) and returns the flight number \ttxt{FLIGHT\_No}, or returns the \ttxt{NO\_FLIGHT} value.
Service 2 works similarly.
The main difference is that, if the flight is not available, it checks whether it is possible to make an overbooking reservation (\ttxt{can\_overbook} flag) and proceeds with the reservation (\ttxt{overbook}) before returning the flight number or the \ttxt{NO\_FLIGHT} value.
Instead, service 3 finds a sequence of flights for the destination, namely an itinerary (\ttxt{generate\_travel\_to}).
Then the itinerary is reserved (\ttxt{reserve}), a travel insurance is stipulated (\ttxt{insurance}) and the itinerary is returned.
Service 4 resembles 3, but no insurance is activated.
Hotel booking services, i.e., services 5 and 6, receive a destination city $x$ and book an hotel (action \ttxt{book}) before returning the hotel reservation \ttxt{HOTEL\_RESV}.
The main difference between the two services is that service 5 looks for a 3 stars hotel (action \ttxt{find\_hotel\_3s}) while service 6, after discriminating on the flag \ttxt{high\_season}, searches either a 2 stars or a 4 stars hotel (actions \ttxt{find\_hotel\_2s} and \ttxt{find\_hotel\_4s}, respectively).
Payment services 7 and 8 receive an item identifier $x$ and return an electronic receipt \ttxt{RCPT} after performing a purchase operation (action \ttxt{buy}).
However, while 8 charges the operation with a constant, extra amount (action \ttxt{const\_charge}), service 7 applies either no commission charge or a variable amount (action \ttxt{var\_charge}).
Finally, signing services accept a document $x$ and return a signed version of it \ttxt{SIGNED\_DOC}.
The only difference between them is that 9 uses a 64 bit key for the signing process (action \ttxt{sign\_64}) while 10 uses a 128 bit ones (\ttxt{sign\_128}).

Note, that with several alternative services which provide the same functionality we have several different possible execution paths which have different security properties.
\end{example}

\begin{example}\label{ex:travel}
We assume the existence of the resources: $\mathcal{A} = \{\ttxt{AIRPORT}\}$ and $\mathcal{C} = \{\ttxt{CITY}\}$.
In Figure~\ref{fig:travel} we propose a \lreq~implementation of the workflow of the BestTravel service, called $e_{B}$.

\begin{figure}
\noindent
\fbox{
\begin{minipage}{0.95\columnwidth}
\footnotesize
\begin{tabular}{ r | l }
 1& $\labs{}{x}{(\gamma\langle\labs{z}{y}{}}$ \\
 2& $\quad\vdots\quad\vdots\quad\ifthen{\ttxt{is\_empty}}{\ttxt{SIGNED\_DOC}}{(\req{\rho_1}{\htype{\mathcal{D}}{}{\mathcal{D}}})y ; z\,y}~\rangle$ \\
 3& $\quad\vdots\quad\forkc \gamma \langle (\labs{}{y'}{(\req{\rho_2}{\htype{\mathcal{B}}{}{\mathcal{D}}})y'}) ((\req{\rho_3}{\htype{\mathcal{C}}{}{\mathcal{H}}})\ttxt{CITY}) \rangle$ \\
 4& $\quad\vdots\quad\andc \gamma \langle \labs{}{y''}{(}\ifc{\ttxt{no\_direct\_flight}}\thenc{(\req{\rho_4}{\htype{\mathcal{B}}{}{\mathcal{D}}})((\req{\rho_5}{\htype{\mathcal{A}}{}{\mathcal{I}}})\,\ttxt{AIRPORT})}$ \\
 5& $\quad\vdots\quad\vdots\quad\elsec{(\req{\rho_6}{\htype{\mathcal{B}}{}{\mathcal{D}}})\,y''})((\req{\rho_7}{\htype{\mathcal{A}}{}{\mathcal{F}}})\,\ttxt{AIRPORT})~\rangle~)$ \\
\end{tabular}
\end{minipage}
}
\caption{Implementation of BestTravel.}
\label{fig:travel}
\end{figure}

%
%

In words, $e_B$ carries out three tasks: it concurrently runs $(i)$ a hotel booking process (line 4) and $(ii)$ a flight booking one (lines 5-6) and, then, $(iii)$ executes a signature procedure (line 2).
The first process consists of an invocation to a hotel search service using the resource \ttxt{CITY}.
The result is then passed as input for (an invocation to) a payment service.
Similarly, the second process requests a itinerary searching service using the resource \ttxt{AIRPORT}.
Then, according the evaluation of the guard \ttxt{no\_direct\_flight}, the service either starts a new request to flight searching service and proceeds with the payment or just invokes a payment service.
The final result of this concurrent execution is the document returned by the first process.
This value is then used as the actual parameter of the last operation of the service.
It consists of a recursive function which, depending on the guard \ttxt{is\_empty}, can either return the resource \ttxt{SIGNED\_DOC} or invoke a signing service and loop.

All the three tasks are subject to a metric requirement $\gamma = \Risk \leq 75$, i.e., each of them must be executed under a risk factor lower than $75$ (\$).
\end{example}

\subsection{C-Semirings}
\label{sec:Semirings}

Our framework exploits the notion of \emph{c-semiring} for the abstraction of metrics and operators over metrics~\cite{BIST-97-JACM}. Usage of this mathematical structure allow us to provide a generic framework for all metrics which could be considered as c-semirings. A c-semiring consists of a set of values $\SemiDom$ (e.g., natural or real numbers), and two types of operators: multiplication ($\otimes$) and summation ($\oplus$) of values and constraints. Formally, a c-semiring is defined as follows (see the work of S. Bistarelli et. al., for more details~\cite{BIST-97-JACM}).

\begin{definition}
A \emph{c-semiring} $\Semiring$ is a tuple $\langle \SemiDom, \oplus, \otimes, \emph{\textbf{0}}, \emph{\textbf{1}}\rangle$ where
\begin{itemize}
\item
\SemiDom~ is a (possibly infinite) set of elements and \emph{\textbf{0}, \textbf{1}} $\in \SemiDom$;
\item
$\oplus$, being an addition defined over $\SemiDom$, is a binary, commutative (i.e., $\semA,\semB \in \SemiDom \Rightarrow \semA \oplus \semB = \semB \oplus \semA$) and associative (i.e., $\semA, \semB, \semC \in \SemiDom \Rightarrow \semA \oplus (\semB \oplus \semC) = (\semA \oplus \semB) \oplus \semC$) operator such that \textbf{0} is its \emph{unit} element (i.e., $\semA \in \SemiDom \Rightarrow (\semA \oplus \emph{\textbf{0}} = \semA = \emph{\textbf{0}} \oplus \semA$);
\item
$\otimes$, being a multiplication over $\SemiDom$, is a binary, commutative and associative operator such that \textbf{1} is its \emph{unit} element and \textbf{0} is its \emph{absorbing} element (i.e., $\semA \in \SemiDom \Rightarrow \semA \otimes \textbf{0} = \textbf{0} = \textbf{0} \otimes \semA$);
\item
$\otimes$ is distributive over additive operator  ($\semA \otimes (\semB \oplus \semC)) = (\semA \otimes \semB) \oplus (\semA \otimes \semC)$;
\end{itemize}
\end{definition}

In this work we focus on a special subset of c-semirings:
\begin{definition}
$c^{*}$-semiring is a c-semiring with $\oplus$ satisfying the following condition: $\forall \semA,\semB \in \SemiDom~\semA\oplus\semB=\semA~or~\semA\oplus\semB=\semB$
\end{definition}

\begin{definition}
$\leq_{\Semiring}$ is a total order over the set $\SemiDom$, such that $\semA \leq_{\Semiring} \semB$ iff $\semA \oplus \semB = \semB$.
\end{definition}

In this work we need a reverse operation for summation $\oplus^{-1}$ which is defined as follows.
\begin{definition}
$\semA \oplus^{-1} \semB = \semA$ iff $\semA \oplus \semB=\semB$.
\label{def:ReverseOPlus}
\end{definition}

In words, this operation always returns the worst possible value.

\begin{property}
Operation $\oplus^{-1}$ is associative, commutative, idempotent, distributive over $\otimes$, and monotone\footnote{A link with proofs:\url{http://www.iit.cnr.it/staff/artsiom.yautsiukhin/Resources/ICE-Proofs.pdf}.}.
\end{property}



\begin{example}
\label{ex:csemirings}
Regarding to the security targets BestTravel is going to use two metrics: trust and risk. Trust is often computed as a probability that the requested service is going to behave as agreed. Thus, trust could be seen as a value between 0 and 1, which is aggregated by multiplying and the higher value is considered better than a lower one. C$^{*}$-semiring for trust value formally is defined as follows: $\langle [0,1], max, \times, 0, 1\rangle$. This type of c-semirings is known as possibilistic semiring.

Risk, considered as possible losses, has the domain of positive real numbers. Multiplication of risks is summation of possible losses, when the lower value is, naturally, considered more preferable than the higher one. Therefore, c$^{*}$-semiring for risk could be seen as $\langle N^{+}\cup\{\infty\}, min, +, \infty, 0\rangle$, known as tropical semiring.
\end{example}

\subsection{Operational Semantics}
\label{sec:OperationalSemantics}

\begin{table}[t]
\centering
\begin{minipage}{\columnwidth}
\hbra
\begin{small}
\begin{tabular}[\textwidth]{r @{$\;\;$} c @{$\quad$} r @{$\;\;$} c}
$\ttxt{(S{-}Ev_1)}$ & $\irule{\osstep{\eta, d, e}{\pi}{\eta', d', e'}}{\osstep{\eta,d, \alpha(e)}{\pi}{\eta',d',\alpha(e')}}$ &
$\ttxt{(S{-}Ev_2)}$ & $\irule{F(\alpha, r) = d'}{\osstep{\eta,d,\alpha(r)}{\pi}{\eta\alpha(r),d \otimes d',\ast}}$ \\ \\
$\ttxt{(S{-}Req)}$ & $\irule{e_{\ell} : \htype{\tau}{H}{\tau'} \in \Srv \qquad \pi(\rho) = {\ell}}{\osstep{\eta,d,(\req{\rho}{\htype{\tau}{}{\tau'}})v}{\pi}{\eta,d,e_{\ell} v}}$ &
$\ttxt{(S{-}App_1)}$ & $\irule{\osstep{\eta,d,e_1}{\pi}{\eta',d',e'_1}}{\osstep{\eta,d,e_1 e_2}{\pi}{\eta',d',e'_1 e_2}}$ \\ \\
$\ttxt{(S{-}App_2)}$ & $\irule{\osstep{\eta,d,e_2}{\pi}{\eta',d',e'_2}}{\osstep{\eta,d,e_1 e_2}{\pi}{\eta',d',e_1 e'_2}}$ &
$\ttxt{(S{-}App_3)}$ & $\osstep{\eta,d,(\labs{z}{x}{e}) v}{\pi}{\eta,d,e\{v/x,\labs{z}{x}{e}/z\}}$ \\ \\
$\ttxt{(S{-}Sec_1)}$ & $\irule{\osstep{\eta,d,e}{\pi}{\eta',d',e'} \quad \eta' \models \varphi}{\osstep{\eta,d,\sfrm{\varphi}{e}}{\pi}{\eta',d',\sfrm{\varphi}{e'}}}$  &
$\ttxt{(S{-}Sec_2)}$ & $\irule{\eta \models \varphi}{\osstep{\eta,d,\sfrm{\varphi}{v}}{\pi}{\eta,d,v}}$ \\ \\
$\ttxt{(S{-}Met_1)}$ & $\irule{\osstep{\eta,d,e}{\pi}{\eta',d',e'} \quad d' \in \gamma}{\osstep{\eta,d,\mfrm{\gamma}{e}}{\pi}{\eta',d',\mfrm{\gamma}{e'}}}$ &
$\ttxt{(S{-}Met_2)}$ & $\irule{d \in \gamma}{\osstep{\eta,d,\mfrm{\gamma}{v}}{\pi}{\eta,d,v}}$ \\ \\
\multicolumn{4}{c}{$\ttxt{(S{-}If)} \quad \osstep{\eta,d,\ifthen{b}{e_\true}{e_\false}}{\pi}{\eta,d,e_{\mathbf{B}(b)}}$}
\end{tabular}
\end{small}
\\
\hket
\end{minipage}
\caption{Operational semantics of \lreq}\label{tab:opsem}
\end{table}

Service execution is driven by the operational semantics defined in Table~\ref{tab:opsem}.
Intuitively, a computation step consists of a reduction from a source configuration to a target one.
Configurations are tuples $\langle \eta, d, e\rangle$ where $\eta$ is an execution trace, i.e., the sequence of events performed so far ($\varepsilon$ denotes the empty execution trace); $d$ is the current metric value; and $e$ is a \lreq~term, which describes the part of the service under evaluation.
The operational semantics is driven by a composition plan $\pi$ which is responsible for providing a mapping between each service request and an actual service, in symbols $\pi(\rho) = \ell$ where $\rho$ and $\ell$ are request and service identifiers, respectively.
In the following we also use $\rightarrow^*_\pi$ for the transitive closure of $\rightarrow_\pi$.

Below, we provide an informal explanation of the operational semantics rules.
To be performed, an action $\alpha$ requires its argument $e$ to be evaluated first (rule \ttxt{(S{-}Ev_1)}).
If the action target reduces to a resource $r$, the action takes place and the current history $\eta$ is extended with the corresponding event $\alpha(r)$ (rule \ttxt{(S{-}Ev_2)}).
Also, the current metric is updated with the metric value for the event $\alpha(r)$.
$F$ is a metric and context-dependent predefined function which assigns a metric value to every event. In practice, function $F$ can be found analytically (e.g., risk=probability$\times$impact), derived form past experience, i.e., using monitoring or assigned by experts (e.g., number of successful virus attacks). A conditional expression is reduced to one of its branches (i.e., $e_\true$ and $e_\false$\footnote{Where $\true$ and $\false$ stand for ``true'' and ``false'', respectively.}) depending on the value of its guard $b$ (rule \ttxt{(S{-}If)}).
Here we assume an evaluation function $\mathbf{B}$, assigning to each possible guard a boolean value, is to be defined.
Rules \ttxt{(S{-}App_1)}, \ttxt{(S{-}App_2)} and \ttxt{(S{-}App_3)} define the behaviour of function application.
Briefly, a function $e$ and its argument $e'$ are both reduced to values, i.e., terms that admit no further reduction.
The steps of the two reductions are executed in a non deterministic way, without any fixed priority between the choice of \ttxt{(S{-}App_1)} and \ttxt{(S{-}App_2)}.
When both computations generate a value, i.e., a lambda abstraction and its argument, the application reduces to the body of the function where the formal parameter $x$ is replaced by the actual value $v$ and the variable $z$ is substituted with the function itself (rule \ttxt{(S{-}App_3)}).
Note that, along the paper, we use $v,v'$ to denote \emph{values}, i.e., closed, effect-free terms being either $\ast$, resources, $\lambda$-abstractions or service requests.
Rules \ttxt{(S{-}Sec_1)} and \ttxt{(S{-}Sec_2)} define the behaviour of the security framing.
Basically, a security framing behaves as its target unless it tries to extend the current history $\eta$ to an illegal trace.
When the target expression reduces to a value, the policy framing can be removed, i.e., the corresponding security check is deactivated, if the current history is a legal one.
Similarly, \ttxt{(S{-}Met_1)} and \ttxt{(S{-}Met_2)} rule metric checks.
In words, a metric check forces metric values generated during the execution of a term $e$ to comply with a constraint $\gamma$.
Finally, service requests (rule \ttxt{(S{-}Req)}) works by running the service $e_\ell$ with actual parameter $v$.
Among all the compatible services, i.e., those having the same behavioural interface specified by the request $\rho$, appearing in the service repository $\Srv$\footnote{Here we assume a service repository to be always available at runtime. In short, a repository is a finite set of tuples, each of them containing at least the service interface and being uniquely identified by the service location $\ell$.}, one is selected according to the current composition plan $\pi$.
Note that the interface of actual services is also annotated with a \emph{history expression} $H$ which represent the service contract (see Section~\ref{sec:types} for more details on this point).

\begin{table}[t]
\begin{center}
\begin{tabular}{c c}
\begin{minipage}{0.45\columnwidth}
\begin{small}
\begin{tabular}[\textwidth]{| c | c | c |}
\hline
{\sc Action} & {\sc Resource} & {\sc Value} \\
\hline
\ttxt{reserve} & \ttxt{FLIGHT\_No} & $15$ \\
\ttxt{reserve} & \ttxt{ITINERARY} & $15$ \\
\hline
\ttxt{overbook} & \ttxt{FLIGHT\_No} & $20$ \\
\hline
\ttxt{insurance} & \ttxt{ITINERARY} & $10$ \\
\hline
\ttxt{find\_hotel\_2s} & \ttxt{CITY} & $30$ \\
\hline
\ttxt{find\_hotel\_3s} & \ttxt{CITY} & $20$ \\
\hline
\ttxt{find\_hotel\_4s} & \ttxt{CITY} & $15$ \\
\hline
\end{tabular}
\end{small}
\end{minipage}
&
\begin{minipage}{0.45\columnwidth}
\begin{small}
\begin{tabular}[\textwidth]{| c | c | c |}
\hline
{\sc Action} & {\sc Resource} & {\sc Value} \\
\hline
\ttxt{book} & \ttxt{HOTEL} & $20$ \\
\hline
\ttxt{var\_charge} & $\cdot$ & $8$ \\
\hline
\ttxt{const\_charge} & $\cdot$ & $5$ \\
\hline
\ttxt{buy} & \ttxt{FLIGHT\_No} & $10$ \\
\ttxt{buy} & \ttxt{HOTEL\_RESV} & $10$ \\
\ttxt{buy} & \ttxt{ITINERARY} & $20$ \\
\hline
\ttxt{sign\_64} & $\cdot$ & $1$ \\
\hline
\end{tabular}
\end{small}
\end{minipage}
\end{tabular}
\end{center}
\caption{Definition of function $F_\Risk$.}\label{tab:riskfun}
\end{table}

\begin{example}\label{ex:execution}

Let $e_1$ be the implementation of service 1 proposed in Example~\ref{ex:serviceimpl}.
We assume $\mathbf{B}(\ttxt{is\_available}) = \true$, and consider the semiring \Risk introduced in Example~\ref{ex:csemirings} and the function $F_\Risk$ which returns the values shown in Table~\ref{tab:riskfun} (where missing entry evaluate to $0$ and $\cdot$ stands for any compatible value).
Then, we have the following computation for $\bigstar = \langle \varepsilon, 0, (e_1)\ttxt{AIRPORT} \rangle$ (where \ttxt{AIRPORT} is a resource in $\mathcal{A}$).

\begin{footnotesize}
$$\bigstar\rightarrow_\pi {\left\langle \varepsilon, 0, \begin{array}{l}\ttxt{search\_flight\_for}(\ttxt{AIRPORT}); \\ \ifc\ttxt{is\_available} \\ \quad\thenc\ttxt{reserve}(\ttxt{FLIGHT\_No}); \\ \qquad \ttxt{FLIGHT\_No} \\ \quad \elsec\ttxt{NO\_FLIGHT} \end{array} \!\!\right\rangle} \rightarrow^*_\pi \left\langle\ttxt{search\_flight\_for}(\ttxt{AIRPORT}), 0, \begin{array}{l} \ifc\ttxt{is\_available} \\ \quad\thenc \\ \;\ttxt{reserve}(\ttxt{FLIGHT\_No}); \\ \; \ttxt{FLIGHT\_No} \\ \quad\elsec\ttxt{NO\_FLIGHT}  \end{array}\!\!\right\rangle$$
$$\rightarrow_\pi\!\!\left\langle\ttxt{search\_flight\_for}(\ttxt{AIRPORT}), 0, \!\!\begin{array}{l}\ttxt{reserve}(\ttxt{FLIGHT\_No});\\\ttxt{FLIGHT\_No}\end{array}\!\!\!\right\rangle \rightarrow^*_\pi\left\langle\begin{array}{c}\ttxt{search\_flight\_for}(\ttxt{AIRPORT})\\\ttxt{reserve}(\ttxt{FLIGHT\_No})\end{array}, 15, \ttxt{FLIGHT\_No}\right\rangle$$
\end{footnotesize}

In words, the computation proceeds as follows.
The first step consists in applying the rule \ttxt{(S{-}App_3)} which, in practice, replaces all the occurrences of $x$ with \ttxt{AIRPORT}.
The second reduction collapses two rules, i.e., \ttxt{(S{-}Ev_2)} and \ttxt{(S{-}App_3)}.
As a result of the rule \ttxt{(S{-}Ev_2)} a new event, that is \ttxt{search\_flight\_for}(\ttxt{AIRPORT}), is added to the execution trace $\varepsilon$.
Also, according to the given definition of $F_\Risk$, the current metric is updated.
Recalling the c$^{*}$-semiring specified in Example~\ref{ex:csemirings}, we note that the multiplication operation over risk values is the sum, then $0 \otimes 0 = 0 + 0 = 0$. 
The subsequent step evaluates the conditional guard \ttxt{is\_available} and chooses the \quote{then} branch (rule \ttxt{(S{-}If)}).
Finally, the last piece of computation repeats the operations described above and updates the current configuration by both adding a new event to the execution history and changing the current metric value (i.e., $0 \otimes 15 = 0 + 15 = 15$).
Since the term appearing in the last configuration is a value, i.e., the resource \ttxt{FLIGHT\_No}, the computation terminates.
\end{example} 

\section{Type and effect system}\label{sec:types}

In this section we present our proposal for a type and effect system.
It derives from the type and effect system presented in~\cite{Bartoletti05csfw} from which it inherits most of its rules.

\subsection{History expressions}\label{subsec:HE}

Briefly, a type and effect system carries out the extraction of behavioural description from a certain expression while typing it.
We use \emph{history expressions} for representing the behaviour of a program in terms of the execution histories it can generate at runtime.

The main novelties introduced by our type and effect system are (i) parallel composition and (ii) metric annotation.
Parallel composition denotes two elements which can run concurrently, in an interleaving fashion.
Instead, metric annotation associate a metric value to a certain behaviour.
Table~\ref{tab:hesyntax} reports the syntax of history expressions.

\begin{table}[t]
\hbra
$\quad H,H' \;\textnormal{::=}\; \varepsilon \orsyn h \orsyn \alpha(r) \orsyn H \cdot H' \orsyn H + H' \orsyn H \mid H' \orsyn \mann{d}{H} \orsyn \sfrm{\varphi}{H} \orsyn \mfrm{\gamma}{H} \orsyn \mu h.H \qquad$
\hket
\caption{Syntax of history expressions}\label{tab:hesyntax}
\end{table}

A history expression can be the empty one $\varepsilon$, a variable $h$ or an access event $\alpha(r)$.
Valid history expressions are also concatenations ($H \cdot H'$), unions ($H + H'$), parallel compositions ($H \mid H'$), metric-annotated expressions ($\mann{d}{H}$), security framings ($\sfrm{\varphi}{H}$), metric checks ($\mfrm{\gamma}{H}$) and least fix-point, recursive expressions ($\mu h.H$).

A history expression denotes a set of execution traces.
We use a denotational semantics to bind each history expression to the corresponding set of traces.
The semantic function $\sem{\cdot}{}{\cdot}$ is defined in Table~\ref{tab:semden}.
Note that we use the environment $\delta$ for mapping variables to set of traces.

\begin{table}[t]
\begin{minipage}{\columnwidth}
\hbra
\vspace{-12pt}
$$\sem{\varepsilon}{\pi}{\delta} = \{ \varepsilon \} \qquad \sem{\alpha(r)}{\pi}{\delta} = \{\alpha(r)\} \qquad \sem{H \cdot H'}{\pi}{\delta} = \sem{H}{\pi}{\delta}\sem{H'}{\pi}{\delta} \qquad \sem{H+H'}{\pi}{\delta} = \sem{H}{\pi}{\delta} \cup \sem{H'}{\pi}{\delta}$$
$$\sem{\sfrm{\varphi}{H}}{\pi}{\delta} = \sfrm{\varphi}{\sem{H}{\pi}{\delta}} \qquad \sem{\mann{d}{H}}{\pi}{\delta} = \sem{H}{\pi}{\delta} \qquad \sem{\mfrm{\gamma}{H}}{\pi}{\delta} = \mfrm{\gamma}{\sem{H}{\pi}{\delta}} \qquad \sem{h}{\pi}{\delta} = \delta(h)$$
$$\sem{H \mid H'}{\pi}{\delta} = \bigcup_{\eta \in \sem{H}{\pi}{\delta} , \eta' \in \sem{H'}{\pi}{\delta}} \binom{\eta}{\eta'} \qquad\qquad \sem{\mu h.H}{\pi}{\delta} = \underset{n > 0}{\bigcup} f^n(\varepsilon) \quad \mathrm{where} \: f(X) = \sem{H}{\pi}{\delta\{X/h\}}$$
where the binary function $\binom{\cdot}{\cdot}$ is recursively defined as follows.
$$(1)\quad\binom{\eta}{\varepsilon} = \{ \eta \} \qquad (2)\quad \binom{\eta}{\alpha\eta'} = \Big\{ \eta_1\alpha\tilde\eta \,\Big|\, \tilde\eta \in \binom{\eta'}{\eta_2} \wedge \eta_1\eta_2 = \eta \Big\}$$
\hket
\end{minipage}
\caption{Denotational semantics}\label{tab:semden}
\end{table}

A $\varepsilon$ expression denotes the singleton containing the empty trace (we use $\varepsilon$ for both void history expressions and empty traces as they are clearly identified by the context).
The semantics of a variable $h$ corresponds to the set of histories associated to it in $\delta$.
A history expression $\alpha(r)$ denotes the singleton $\{\alpha(r)\}$.
The semantics of a sequence $H\cdot H'$ is the set of traces $\eta\eta'$ such that $\eta \in \sem{H}{}{\delta}$ and $\eta' \in \sem{H'}{}{\delta}$.
Similarly, the semantics of a choice is the union between the sets denoted by the two sub-expressions.
Parallel history expressions $H \mid H'$ denote the set of all the possible interleaving of traces belonging to the two sub-expressions.
Interleaving semantics is defined through the binary operator $\binom{\cdot}{\cdot}$.
Intuitively, if one of the two considered histories is $\varepsilon$, the operator $\binom{\cdot}{\cdot}$ returns the other one.
Instead, for non-empty traces it generates all the possible sequences representing concurrent executions.
This process is obtained by considering all the possible prefixes of one trace, adding the first action of the other trace and recursively applying the $\binom{\cdot}{\cdot}$ operator to the remaining \quote{tails}.
In the style of~\cite{Bartoletti07fossacs}, security framing denotes execution histories wrapped between two special actions $[_\varphi$ and $]_\varphi$ (for brevity, we write $\sfrm{\varphi}{X}$ in place of $[_\varphi \cdot X \cdot ]_\varphi$).
These special actions mark the activation and deactivation points of a policy.
Following a similar reasoning, the semantics of $\mfrm{\gamma}{H}$ is the set of traces denoted by $H$ wrapped by the special actions $\langle_\gamma$ and $\rangle_\gamma$ (with the obvious meaning).
Finally, $\mu h.H$ denotes a fix point operation over the set of traces denoted by $H$ (see~\cite{Bartoletti07fossacs} for further detail).

Moreover, we introduce a partial order relation $\sqsubseteq$ between history expressions such that $H \sqsubseteq H' \Leftrightarrow \forall \delta . \sem{H}{}{\delta} \subseteq \sem{H'}{}{\delta}$.


\subsection{Typing relation}

In the following we introduce our typing rules.
The main difference with respect to the rules proposed in previous works is that here we generate metric annotated history expressions during the typing process.
Before presenting the typing rules, we need to introduce \emph{types} and \emph{type environments}.

\begin{definition}{(Types and type environments)}
\\ [1.5ex]
\begin{minipage}{\columnwidth}
\hbra
\begin{center}
$\tau,\tau' \:\mathrm{::=}\: \unit \mid \mathcal{R} \mid \htype{\tau}{H}{\tau'} \qquad\qquad \Gamma,\Gamma' \:\mathrm{::=}\: \emptyset \mid \Gamma;x : \tau$
\end{center}
\hket
\end{minipage}
\end{definition}

A type can be both a simple type, i.e., $\unit$ or the resource domain $\mathcal{R}$\footnote{For simplicity here we assume a single set $\mathcal{R}$, but, in general, we assume to have a finite number of resource domains $\mathcal{R}_1, \ldots, \mathcal{R}_n$ s.t. $\bigcup_i \mathcal{R}_i = \mathcal{R}$}, or a function from type $\tau$ to type $\tau'$.
Functional types also carry a history expression $H$ which represents the latent effect of invoking the function.
Then, a type environment $\Gamma$, being either the empty one $\emptyset$ or the one obtained through a new binding $\Gamma;x : \tau$, is a mapping from variables to types.

The typing relation has the form $\trel{\Gamma, H}{}{e : \tau}$.
It must be read as \quote{under the environment $\Gamma$ and carrying the effect $H$, expression $e$ has type $\tau$}.
The rules in Table~\ref{tab:typerel} define the typing relation.

\begin{table}[t]
\centering
\begin{minipage}{\columnwidth}
\hbra
\vspace{6pt}
\begin{footnotesize}
\begin{tabular}[\textwidth]{c}
$\ttxt{(T{-}Unit)}\;\trel{\Gamma,\varepsilon}{}{\ast : \unit} \quad \ttxt{(T{-}Res)}\;\trel{\Gamma,\varepsilon}{}{r : \mathcal{R}} \quad \ttxt{(T{-}Var)}\;\trel{\Gamma,\varepsilon}{}{x : \Gamma(x)} \quad \ttxt{(T{-}Abs)}\;\irule{\trel{\Gamma;x:\tau;z:\htype{\tau}{H}{\tau'},H}{}{e : \tau'}}{\trel{\Gamma,\varepsilon}{}{\labs{z}{x}{e}:\htype{\tau}{H}{\tau'}}}$ \\ \\
$\ttxt{(T{-}Ev)}\;\irule{\trel{\Gamma,H}{}{e : \mathcal{R}}}{\trel{\Gamma,H\cdot\underset{r \in \mathcal{R}}{\sum}(\mann{F(\alpha,r)}{\alpha(r)})}{}{\alpha(e) : \unit}} \quad \ttxt{(T{-}App)}\;\irule{\trel{\Gamma,H}{}{e : \htype{\tau}{H''}{\tau'}} \quad \trel{\Gamma,H'}{}{e' : \tau}}{\trel{\Gamma,(H \mid H') \cdot H''}{}{e\,e' : \tau'}} \quad \ttxt{(T{-}Frm)}\;\irule{\trel{\Gamma,H}{}{e : \tau}}{\trel{\Gamma,\sfrm{\varphi}{H}}{}{\sfrm{\varphi}{e} : \tau}}$ \\ \\
$\ttxt{(T{-}Met)}\,\irule{\trel{\Gamma,H}{}{e : \tau}}{\trel{\Gamma,\mfrm{\gamma}{H}}{}{\mfrm{\gamma}{e} : \tau}} \quad \ttxt{(T{-}If)}\;\irule{\trel{\Gamma,H}{}{e : \tau} \quad \trel{\Gamma,H}{}{e' : \tau}}{\trel{\Gamma,H}{}{\ifthen{g'}{e}{e'} : \tau}} \quad \ttxt{(T{-}Wkn)}\;\irule{\trel{\Gamma,H}{}{e : \tau} \quad H \sqsubseteq H'}{\trel{\Gamma,H'}{}{e : \tau}}$ \\ \\
$\ttxt{(T{-}Req)}\;\irule{I = \{H \mid e_{\ell} : \htype{\tau}{H}{\tau'} \in \Srv\}}{\trel{\Gamma,\varepsilon}{}{\req{\rho}{\htype{\tau}{}{\tau'}}} : \htype{\tau}{\sum_{X \in I} X}{\tau'}}$
\\ [3.0ex]
\end{tabular}
\end{footnotesize}
\hket
\end{minipage}
\caption{Typing relation}\label{tab:typerel}
\end{table}

Briefly, the expression $\ast$ has $\unit$ type and generates no side effects ($H = \varepsilon$, rule \ttxt{(T{-}Unit)}) while a resource $r$, being also side effect free, has type $\mathcal{R}$ (rule \ttxt{(T{-}Res)}).
The type of a variable $x$ depends on the typing context provided by $\Gamma$ (rule \ttxt{(T{-}Var)}).
Abstractions (rule \ttxt{(T{-}Abs)}) has an empty effect and produce a functional type \htype{\tau}{H}{\tau'} from their input to their output types.
The latent effect $H$ is the one obtained from typing the function body.
Rule \ttxt{(T{-}Ev)} requires more attention.
Indeed, we say that an expression $\alpha(e)$, having type $\unit$, generates a history expression which is the sequence between the history expression deriving from typing its argument $e$ and the summation (i.e., a finite sequence of choice operators) of all the possible access actions $\alpha$ to a compatible resource $r \in \mathcal{R}$.
Also, all of these access events are annotated with the metric value provided by the function $F$.
The application of a function $e$ to an argument $e'$, i.e., rule \ttxt{(T{-}App)}, has type equal to the return type of $e$ and a history effect which is the sequence between (1) the two effects of $e$ and $e'$ in parallel and (2) the latent effect of the function.
Security and metric framing (rules \ttxt{(T{-}Frm)} and \ttxt{(T{-}Met)}) have the same type as their targets and produce wrapped history expressions.
Rule \ttxt{(T{-}Wkn)} says that we can always type an expression under a more general history expression.
Finally, rule \ttxt{(T{-}Req)} says that a service request has the same type of its signature but for its latent effect which is obtained as the disjunction of all the (latent effects of the) possible servers appearing in the repository $\Srv$.

\begin{example}\label{ex:typing}
Consider service 9, we call its implementation $e_{9}$, of Example~\ref{ex:serviceimpl}.
Writing it without abbreviations we obtain: $e_{9} = \labs{z}{x}{(\labs{w}{y}{\ttxt{SIGNED\_DOC})\ttxt{sign\_64}(x)}}$.
Then consider the function $F_\Risk$ of Table~\ref{tab:riskfun}.
We type $e_9$ as follows.
\begin{small}
\[
\irule{
 \irule{
  \irule{\trel{\Gamma', \varepsilon}{}{\ttxt{SIGNED\_DOC} : \mathcal{D}}}
  {\trel{\Gamma, \varepsilon}{}{\labs{w}{y}{\ttxt{SIGNED\_DOC}} : \htype{\tau}{\varepsilon}{\mathcal{D}}}}
  \quad
  \irule{\Gamma(x) = \mathcal{D}}{\trel{\Gamma, H_9}{}{\ttxt{sign\_64}(x) : \unit}}
 }
 {
  \trel{\Gamma, H_9}{}{(\labs{w}{y}{\ttxt{SIGNED\_DOC})\ttxt{sign\_64}(x) : \mathcal{D}}}
 }
}
{\trel{\emptyset, \varepsilon}{}{e_9 : \htype{\mathcal{D}}{H_9}{\mathcal{D}}}}
\]
\end{small}

\noindent
where $H_9 = {\mann{1}{\ttxt{sign\_64}(\ttxt{RCPT})} + \mann{1}{\ttxt{sign\_64}(\ttxt{SIGNED\_DOC})}}$, $\Gamma = x : \mathcal{D}; z : \htype{\mathcal{D}}{H_9}{\mathcal{D}}$ and $\Gamma' = \Gamma; y : \tau; w : \htype{\tau}{\varepsilon}{\mathcal{D}}$.
Following a similar reasoning we type all the services of example~\ref{ex:serviceimpl} as shown in Figure~\ref{fig:typing}.
For brevity, in the following we use $H_i$ to denote the latent effect of service $e_i$.
\end{example}

\begin{figure}
\centering
\begin{tabular}{@{$\!$} r @{\hspace{2pt}} l}
$e_1 :$ & $\htype{\mathcal{A}}{\footnotesize
\begin{array}{c}
\mann{0}{\ttxt{search\_flight\_for}(\ttxt{AIRPORT})} \cdot \\ \left(\left(\begin{array}{c}
\mann{15}{\ttxt{reserve}(\ttxt{FLIGHT\_No})} +\\
\mann{0}{\ttxt{reserve}(\ttxt{NO\_FLIGHT})}
\end{array}\right) +
\left(\begin{array}{c}
\mann{20}{\ttxt{overbook}(\ttxt{FLIGHT\_No})} +\\
\mann{0}{\ttxt{overbook}(\ttxt{NO\_FLIGHT})} + \varepsilon
\end{array}\right)\right)
\end{array}
}{\mathcal{F}} $ \\
$e_2 :$ & $\htype{\mathcal{A}}{\footnotesize
(\mann{0}{\ttxt{search\_flight\_for}(\ttxt{AIRPORT})}) \cdot (\mann{15}{\ttxt{reserve}(\ttxt{FLIGHT\_No})} +
\mann{0}{\ttxt{reserve}(\ttxt{NO\_FLIGHT})})
}{\mathcal{F}} $ \\
$e_3 :$ & $\htype{\mathcal{A}}{\footnotesize (\mann{0}{\ttxt{generate\_travel\_to}(\ttxt{AIRPORT})})
\cdot (\mann{15}{\ttxt{reserve}(\ttxt{ITINERARY})}) \cdot (\mann{10}{\ttxt{insurance}(\ttxt{ITINERARY})})
}{\mathcal{I}} $ \\
$e_4 :$ & $\htype{\mathcal{A}}{\footnotesize
(\mann{0}{\ttxt{generate\_travel\_to}(\ttxt{AIRPORT})})
\cdot (\mann{15}{\ttxt{reserve}(\ttxt{ITINERARY})})
}{\mathcal{I}} $ \\
$e_5 :$ & $\htype{\mathcal{C}}{\footnotesize
(\mann{20}{\ttxt{find\_hotel\_3s}(\ttxt{CITY})})
\cdot (\mann{20}{\ttxt{book}(\ttxt{HOTEL})})
}{\mathcal{H}} $ \\
$e_6 :$ & $\htype{\mathcal{C}}{\footnotesize
(\mann{30}{\ttxt{find\_hotel\_2s}(\ttxt{CITY})} + \mann{15}{\ttxt{find\_hotel\_4s}(\ttxt{CITY})}) \cdot
(\mann{20}{\ttxt{book}(\ttxt{HOTEL})})
}{\mathcal{H}} $ \\
$e_7 :$ & $\htype{\mathcal{B}}{\footnotesize
\varepsilon + \left(\left(\begin{array}{c}
\mann{8}{\ttxt{var\_charge}(\ttxt{ITINERARY})} + \\ \mann{8}{\ttxt{var\_charge}(\ttxt{FLIGHT\_No})} + \\
\mann{8}{\ttxt{var\_charge}(\ttxt{NO\_FLIGHT})} + \\
\mann{8}{\ttxt{var\_charge}(\ttxt{HOTEL\_RESV})} \end{array}\right)
\cdot
\left(\begin{array}{c}
\mann{20}{\ttxt{buy}(\ttxt{ITINERARY})} + \\
\mann{10}{\ttxt{buy}(\ttxt{FLIGHT\_No})} + \\
\mann{0}{\ttxt{buy}(\ttxt{NO\_FLIGHT})} + \\
\mann{10}{\ttxt{buy}(\ttxt{HOTEL\_RESV})}
\end{array}\right)\right)}{\mathcal{D}} $ \\
$e_8 :$ & $\htype{\mathcal{B}}{\footnotesize
\left(\begin{array}{c}
\mann{5}{\ttxt{const\_charge}(\ttxt{ITINERARY})} + \\ \mann{5}{\ttxt{const\_charge}(\ttxt{FLIGHT\_No})} + \\
\mann{5}{\ttxt{const\_charge}(\ttxt{NO\_FLIGHT})} + \\
\mann{5}{\ttxt{const\_charge}(\ttxt{HOTEL\_RESV})} \end{array}\right)
\cdot
\left(\begin{array}{c}
\mann{20}{\ttxt{buy}(\ttxt{ITINERARY})} + \\
\mann{10}{\ttxt{buy}(\ttxt{FLIGHT\_No})} + \\
\mann{0}{\ttxt{buy}(\ttxt{NO\_FLIGHT})} + \\
\mann{10}{\ttxt{buy}(\ttxt{HOTEL\_RESV})}
\end{array}\right)}{\mathcal{D}} $ \\
$e_9 :$ & $\htype{\mathcal{D}}{\footnotesize \mann{1}{\ttxt{sign\_64}(\ttxt{RCPT})} + \mann{1}{\ttxt{sign\_64}(\ttxt{SIGNED\_DOC})}}{\mathcal{D}} $ \\
$e_{10} :$ & $\htype{\mathcal{D}}{\footnotesize \mann{0}{\ttxt{sign\_128}(\ttxt{RCPT})} + \mann{0}{\ttxt{sign\_128}(\ttxt{SIGNED\_DOC})}}{\mathcal{D}} $ \\
\end{tabular}
\caption{Types inferred from the services of Example~\ref{ex:serviceimpl}.}\label{fig:typing}
\end{figure}

\begin{example}\label{ex:bttype}
Using the notation introduced in the previous examples for denoting the history expressions of services, we type the BestTravel implementation $e_B$ as in Figure~\ref{fig:bttype}.
We call $H_B$ the latent effect labelling the arrow type of $e_B$.
\end{example}

\begin{figure}[th]
\[
e_B : \htype{\unit}{\footnotesize
\left(
	\left.
		\mfrm{\gamma}{(H_1 + H_2) \cdot \left(\begin{array}{c}(H_7 + H_8) \\ + \\ ((H_3 + H_4) \cdot (H_7 + H_8))\end{array}\right)}
	\right\vert
	\mfrm{\gamma}{\begin{array}{c}
		(H_5 + H_6)\\
		\cdot \\
		(H_7 + H_8)
	\end{array}}
\right) \cdot \mfrm{\gamma}{\mu h.((H_9 + H_{10}) \cdot h + \varepsilon)}}{\mathcal{D}}
\]
\caption{Type of BestTravel.}\label{fig:bttype}
\end{figure}

The main result on the type and effect system is \emph{type safety}.
In words, type safety guarantees that effects produced by the type and effect system safely denote the behaviour of services.

\begin{theorem}\label{thm:typesaf}
If $\trel{\Gamma,H}{true}{e : \tau}$ and $\langle\varepsilon, d, e\rangle \rightarrow^\ast_\pi \langle\eta,d',v\rangle$ then $\forall\delta.\eta \in \sem{H}{\sigma}{\delta}$.
\end{theorem}

Interestingly, the extensions presented in this paper do not invalidate this result originally proved by Bartoletti et. al.~\cite{Bartoletti05csfw}.
In the next section, we show that history expressions safety is also preserved under metric factorization.

\section{Security and metric analysis}
\label{sec:analysis}

\subsection{History expressions and semirings}

Metric annotations are used to label a history expression with metric values which are expected to be produced dynamically.
However, metric annotations are locally associated with parts of a history expression while, in general, it would be preferable to have a single value labelling the whole expression.
In particular, we are interested in a procedure which turns a history expression into a corresponding normal form. 

\begin{definition}
A history expression $H$ is said to be in \emph{metric normal form} (MNF), iff $H = \mann{d}{H'}$ and $H'$ contains no metric annotations.
\end{definition}

In Table~\ref{tab:eqthy} we propose a set of equivalences that we use to move and compose metric annotations appearing in history expressions. The rules in Table~\ref{tab:eqthy} define the correspondence between the history expressions and the semiring operators.
In particular, we can always add a multiplication-neutral annotation to a history expression, nested annotations are commutative and can be reduced to a semiring multiplication and choice correspond to the inverse of a semiring addition, namely a subtraction.
Also parallel composition can be annotated with the (result of the) multiplication between the two subexpressions annotations.
A security framing is orthogonal to metric annotation, i.e., they do not affect each other.
Instead, metric checks have a precise effect on annotations.
As a matter of fact, we can remove a metric check by forcing its target to be annotated with the difference ($\oplus^{-1}$) between the inner annotation and the threshold of $\gamma$.
Finally, a recursion is annotated with the least fix point of the function $\Phi$ that extracts the metric annotation from the inner history expression after annotating the bounded variable $h$.

\begin{table}[t]
\centering
\begin{minipage}{\columnwidth}
\hbra
\vspace{-12pt}
$$H \equiv \mann{\mathbf{1}}{H} \quad \mann{d_1}{\mann{d_2}{H}} \equiv \mann{d_2}{\mann{d_1}{H}} \equiv \mann{d_1 \otimes d_2}{H} \quad \mann{d_1}{H_1} \cdot \mann{d_2}{H_2} \equiv \mann{d_1 \otimes d_2}{(H_1 \cdot H_2)}$$
$$\mann{d_1}{H_1} + \mann{d_2}{H_2} \equiv \mann{d_1 \oplus^{-1} d_2}{(H_1 + H_2)} \quad \mann{d_1}{H_1} \mid \mann{d_2}{H_2} \equiv \mann{d_1 \otimes d_2}{(H_1 \mid H_2)} \quad \sfrm{\varphi}{\mann{d}{H}} \equiv \mann{d}{\sfrm{\varphi}{H}}$$
$$\mfrm{\gamma}{\mann{d}{H}} \equiv \mann{\bar d}{\mfrm{\gamma}{H}} \quad \textnormal{where }\gamma = T \geq_T d' \textnormal{ and } \bar d = d \oplus^{-1} d'$$
$$\mu h.H \equiv \mann{\bar{d}}{\mu h.H'} \qquad \textnormal{where} \quad \bar{d} = \bigoplus_n{}^{-1} \Phi^n(\mathbf{0}) \textnormal{ and } \Phi(d) = d' \Leftrightarrow \left\lbrace\begin{array}{c} H[\bind{h}{\mann{d}{h}}] \equiv \mann{d'}{H'} \\ \wedge \\ \mann{d'}{H'} \textnormal{ is in MNF} \end{array}\right.$$
\hket
\end{minipage}
\caption{Equational rules.}\label{tab:eqthy}
\end{table}

A crucial property we want to prove on the equation rules of Table~\ref{tab:eqthy} is that they do not invalidate the semantics of history expressions.
Such property guarantees that history expression transformations do not affect the safety property stated by theorem~\ref{thm:typesaf}.

\begin{property}\label{prop:equivsafe}
For all history expressions $H$ and $H'$ if $H \equiv H'$ then $\forall \delta.\sem{H}{}{\delta} = \sem{H'}{}{\delta}$
\end{property}

\begin{example}\label{ex:mnf}
Having in mind that $\oplus^{-1}$ is \emph{max} for \Risk, consider the history expression $H_2$ of Example~\ref{ex:typing}
\begin{align}
&H_2 = (\mann{0}{\ttxt{search\_flight\_for}(\ttxt{AIRPORT})}) \cdot (\mann{0}{\ttxt{reserve}(\ttxt{FLIGHT\_No})} + \mann{15}{\ttxt{reserve}(\ttxt{NO\_FLIGHT})}) \nonumber \\
&H_2 \equiv \mann{0 \otimes (0 \oplus^{-1} 15)}{(\ttxt{search\_flight\_for}(\ttxt{AIRPORT})} \cdot (\ttxt{reserve}(\ttxt{FLIGHT\_No}) + \ttxt{reserve}(\ttxt{NO\_FLIGHT})))\nonumber
\end{align}

Note, that the right side of the previous equivalence is in MNF.
According to the operations of the semiring \Risk, the resulting annotation value is $15$.
\end{example}

\begin{example}\label{ex:mnfcnt}
We write the MNF of the history expressions of Example~\ref{ex:typing}.
For brevity, we write $H_i \equiv \mann{d_i}{H'_i}$ to emphasise the metric annotation of the MNF without showing the structure of $H'_i$.
\begin{center}
\begin{tabular}{r @{$\;\equiv\;$} l @{$\;\qquad\;$} r @{$\;\equiv\;$} l @{$\;\qquad\;$} r @{$\;\equiv\;$} l @{$\;\qquad\;$} r @{$\;\equiv\;$} l @{$\;\qquad\;$} r @{$\;\equiv\;$} l}
$H_1$ & $\mann{20}{H'_1}$ & $H_2$ & $\mann{15}{H'_2}$ & $H_3$ & $\mann{25}{H'_3}$ & $H_4$ & $\mann{15}{H'_4}$ & $H_5$ & $\mann{40}{H'_5}$ \\
$H_6$ & $\mann{50}{H'_6}$ & $H_7$ & $\mann{28}{H'_7}$ & $H_8$ & $\mann{25}{H'_8}$ & $H_9$ & $\mann{1}{H'_9}$ & $H_{10}$ & $\mann{0}{H'_{10}}$ \\
\end{tabular}
\end{center}
\end{example}

Intuitively, Example~\ref{ex:mnfcnt} shows that every history expression appearing in our working example has an equivalent MNF.
In general, we know that all the history expressions can be reduced to a corresponding MNF as stated by the following property.

\begin{property}
For each history expression $H$ there exists $H'$ such that $H \equiv H'$ and $H'$ is in MNF.
\end{property}

The last property we show is \emph{metric safety}, which characterizes the most important quality of the metric annotations we generate.

\begin{theorem}
If $\trel{\Gamma, H}{}{e : \tau}$ and $H \equiv \mann{\bar d}{H'}$ such that $\mann{\bar d}{H'}$ is in MNF, then for each execution $\tosstep{\eta,d,e}{\pi}{\eta',d',e'}$ holds that $d' \leq_T d \otimes \bar d$.
\end{theorem}

Similarly to type safety, this theorem guarantees that metric annotations produced by our equational theory provide an upper bound to the metric values generated by the execution of a term.
As each of them has a corresponding MNF, this theorem can be universally applied to any history expression.
%
%

\subsection{Discussion}

During the presentation we have shown how our formalism can be applied to the modelling of complex business processes.
In this part of the article we describe how the proposed theory can be applied to the verification and analysis of the security properties of web services.

Basically, our proposal offers facilities that can be applied to all the stages of service design, implementation and execution.
Statically, service designers can write their policies on execution histories and security metrics.
Then, developers apply the scope of the policies to the service implementation.
Finally, each service runs with proper checks controlling that the execution complies with the specification.

These steps suffice to carry out the analysis of possible configurations of a complex abstract business process.
The goal is to check whether the possible configurations satisfy desired policies.
This information is required in order to decide if we can avoid run-time controls.
Naturally, if a configuration satisfies a policy or the worst possible metric value is better than a threshold, there is no need for an additional control. 

Note, that we assume that the declared policies/metrics for a specific service are genuine and the services are typed by a trusted type and effect system (implementation).
Although this assumption is not true in general, here we focussed on the considered problem, i.e., aggregation of metrics and check of composite properties.

In order to check that a certain business process satisfies properties or has sufficiently good metric value the analyst starts for a \lreq~implementation
of an abstract workflow, as it is shown in Example~\ref{ex:travel}.
Then, we assume the service repository and c$^{*}$-semirings for considered metrics to be defined similar to Examples~\ref{ex:serviceimpl} and~\ref{ex:csemirings}.
The next step is to type the service implementation similar to Example~\ref{ex:typing} and~\ref{ex:bttype}.
Finally, we aggregate metrics annotations, as it is done in Example~\ref{ex:mnf}.
During this process, several analysis on the validity of history expressions can be carried out in order to prevent illegal service compositions.
For a description of these techniques we refer the interested reader to~\cite{Bartoletti05csfw, Costa10modular}.




\begin{example}
We use the history expressions in MNF shown in Example~\ref{ex:mnfcnt} to compute the MNF of $H_B$.
Considering the history expression appearing in Figure~\ref{fig:bttype}, we can replace every instance of $H_i$ with the corresponding MNF $\mann{d_i}{H'_i}$.
Then we obtain the following equivalences.
\[
H_B \equiv (\mfrm{\gamma}{H_F} \mid \mfrm{\gamma}{H_H}) \cdot \mfrm{\gamma}{H_S}
\]
\[
H_F \equiv \Big( (\mann{20}{H'_1} + \mann{15}{H'_2}) \cdot \big( (\mann{28}{H'_7} + \mann{25}{H'_8}) + ((\mann{25}{H'_3} + \mann{15}{H'_4}) \cdot (\mann{28}{H'_7} + \mann{25}{H'_8})) \big) \Big)
\]
\[
H_H \equiv \Big( (\mann{40}{H'_5} + \mann{50}{H'_6}) \cdot (\mann{28}{H'_7} + \mann{25}{H'_8}) \Big)
\qquad
H_S \equiv \mu h.((\mann{1}{H'_9} + \mann{0}{H'_{10}}) \cdot h + \varepsilon)
\]
Applying the rules of Table~\ref{tab:eqthy}, we can reduce to the following history expression.
\[
H_B \equiv (\mfrm{\gamma}{\mann{73}{H'_F}} \mid \mfrm{\gamma}{\mann{78}{H'_H}}) \cdot \mfrm{\gamma}{\mann{\infty}{H'_S}}
\]
Recalling that $\gamma = \Risk \leq 75$ we conclude with the equivalences below.
$$(\mfrm{\gamma}{\mann{73}{H'_F}} \mid \mfrm{\gamma}{\mann{78}{H'_H}}) \cdot \mfrm{\gamma}{\mann{\infty}{H'_S}} \equiv (\mann{73}{\mfrm{\gamma}{H'_F}} \mid \mann{75}{\mfrm{\gamma}{H'_H}}) \cdot \mann{75}{\mfrm{\gamma}{H'_S}} \equiv \mann{223}{((\mfrm{\gamma}{H'_F} \mid \mfrm{\gamma}{H'_H}) \cdot \mfrm{\gamma}{H'_S})}$$

Interestingly, we note that, among the three instances of $\gamma$, only the first one applies to a history expression
satisfying the restriction, i.e., $73 \in \gamma$.
We cannot say the same for the other two instances.
However, our semantics for metric framing forces the execution of all the parts of the service to respect risk constraints.
In this way, even though some parts of the service are labelled with $\infty$, the overall risk is a finite value, i.e., $223$.

Since the last two instances fail the restriction the dynamic analysis is required. Note, that the hotel reservation part of the process may use services $H_{5}$ and $H_{8}$ with the overall risk level $65<75$. Therefore, during the execution we guard the second and the third instances to guarantee the low risk level values. There is no need to guard the first instance, since it satisfies the restriction in any case. Imagine, that during the execution $H_{6}$ service has been selected. Before executing the next step the guard must check the resulting value, using the same rules as for the static analysis. In case $H_{8}$ is selected the execution is allowed ($75\leq75$). Otherwise, if $H_{7}$ is chosen the restriction fails ($78>75$) and the execution is halted (or another action is performed,
e.g., a report about the failure is sent to the customer and provider).

%

%
%
%
\end{example}

\section{Related work}\label{sec:related}

Outsourcing processing of sensitive data to external parties requires some assurances, that the data will be well protected while processed and transmitted. Unsurprisingly, several authors claimed that security requirements must be included into the agreement between service customer and service provider \cite{IRVI-00-NSPW,KARJ-05-QoP}. Our work extends the existing state of the art with a unified approach for checking security properties and security metrics of complex business processes which appear as statements in such agreements.

Many authors proposed formal languages for specifying and verifying agreements, also called \emph{contracts}, between a service provider and a customer.
Padovani~\cite{Padovani08contract} proposes a language for defining service contracts and presents a theory for the automatic generation of service orchestrators.
Subcontract relations are used to find a matching between the contract offered by a service and the requirements of its clients.
Similarly, Bravetti and Zavattaro~\cite{Bravetti07towardsa} present a language for the specification of service contracts.
Their contracts have a process algebra-based semantics and allow for the specification of composed services.
Contract composition can be verified to guarantee that the interaction of a group of services does not violate the specifications.
Even though these works do not focus on security analysis, their contracts can be adapted to model security requirements.

Martinelli and Matteucci~\cite{Martinelli07synthesis} presented a framework for the synthesis of a secure orchestrator, i.e., an agent which drives the interaction between two services guaranteeing that  a certain security policy is respected.
Although, the proposals described above use contracts for the specification and analysis of history-based~\cite{Abadi03access} service properties, none of them allows for the definition of security metrics and restrictions on them.

In order to check whether a complex business process satisfies some quantitative requirements aggregation of security metric values for atomic services is required. For example, Cheng et. al. \cite{CHEN-05-TECHREP} aggregated downtime metric, considering business process like a simple set of activities, i.e., regardless the operational flow.


In contrast, Jaeger et. al. \cite{JEAG-05-EEE} have shown that some metrics could be aggregated differently depending on the structural activity used for joining the atomic services. In this work all metrics were considered separately. Moreover, the author did not considered security metrics. Yu et. al. \cite{YU-07-ACMTRANS,YU-05-EEE} applied the idea of Jeager at. al. for selection of the best process among several alternatives. The authors defined aggregation functions for several metrics and aimed at selection of the best alternative which satisfies the constraints specified in the agreement. First the authors defined a utility function and proposed to solve a 0-1 multi-dimension multi-choice knapsack problem (MMKP) only for a sequential order \cite{YU-05-EEE}. Solutions for a general workflow were proposed later \cite{YU-07-ACMTRANS}.


Massacci and Yautsiukhin \cite{MASS-07-QoP} proposed a method and an algorithm for aggregation of security metrics. The authors also solved the problem of selection the best (i.e., more secure) alternative, though a wider range of metrics were considered (these metrics cannot be used in classical algorithms for finding the shortest path). The method was extended for checking several metrics at the same time using Pareto optimality strategy \cite{INNE-08-NordSec}.

In our work we do not have a goal to select the business process which has the best metric value. Moreover, we assume that some processes which do have a value worse than desired may still satisfy the policy if a more secure execution path is selected for a specific invocation. 
Therefore, our proposal allows making a decision at design time and supporting control at run-time. 

\section{Conclusion}\label{sec:conclusion}

In this paper we presented a novel approach for dealing with the analysis and verification of both security and metric requirements of web services.
Our system is developed on existing solutions for modelling security and metric-based requirements.
The result is a unified framework for ($i$) the definition and application of security and metric policies within service implementation, ($ii$) the automatic extraction of history expressions carrying metric annotations and ($iii$) the computation (through an equational theory) of  metric values which safely predict the expected behaviour of services.
Our proposal requested a new type and effect system, extending existing approaches, to be defined.
Interestingly, we found that adding metric annotations does not invalidate the type safety property, i.e., annotations are orthogonal to the history expressions.

The present work is a first step toward a complete model for the specification and verification of quantitative and qualitative, non functional requirements for web services.
Further effort is requested in order to generalise our approach.
In particular, we aim at defining a procedure for generating orchestration plans starting from the history expressions produced by our type and effect system.
Such method has been presented in~\cite{Bartoletti05csfw} for metric-free history expressions and we believe that similar results can be extended to our proposal.
Another limitation of the current model is our static description of metric value for the events.
Even though we think that assigning metric values to events is a reasonable way to model the actual behaviour of services, it is not always correct to assume these values to keep unchanged in time.
Indeed, many metrics aim at modelling dynamic evolution of some property, e.g., reputation or number of system failures, which we cannot model with our approach.

%


%
\begin{appendix}

\end{appendix}

\end{document}